\documentclass[pdftex,twocolumn,epjc3]{svjour3}          

\RequirePackage[T1]{fontenc}

\smartqed  

\RequirePackage{graphicx}
\RequirePackage{mathptmx}      
\RequirePackage{flushend}
\RequirePackage[numbers,sort&compress]{natbib}
\RequirePackage[colorlinks,citecolor=blue,urlcolor=blue,linkcolor=blue]{hyperref}

\usepackage{graphicx,epstopdf}
\usepackage{dcolumn}
\usepackage{bm}

\newcommand{\nn}{\nonumber}

\journalname{Eur. Phys. J. C}

\begin{document}

\title{Null Gravitational Redshift by a Reissner-Nordstr\"{o}m Black Hole in the Strong Field Limit}

\author{Guansheng He\thanksref{addr1,e1}
        \and
        Chaohong Pan\thanksref{addr1}
        \and
        Xia Zhou\thanksref{addr2}
        \and
        Weijun Li\thanksref{addr1}
        \and
        Lin Li\thanksref{addr1,e2}
        }

\thankstext{e1}{e-mail: hgs@usc.edu.cn (corresponding author)}
\thankstext{e2}{e-mail: lilinmath@usc.edu.cn}

\institute{School of Mathematics and Physics, University of South China, Hengyang, 421001, China\label{addr1}
          \and
          Physics and Space Science College, China West Normal University, Nanchong, 637009, China\label{addr2}
}

\date{Received: date / Accepted: date}

\maketitle

\begin{abstract}
The gravitational shift of electromagnetic frequency in the strong field limit is usually investigated under the common scenario, where the light receiver is far away from the central body while the emitter is in the strong-field region of the lens. In this paper, the gravitational frequency shift of light caused by a Reissner-Nordstr\"{o}m (RN) black hole is studied numerically in the traditional strong-field scenario, as well as in the scenario where both the light emission and reception events happen in the strong-field region of the black hole. In order to obtain the numerical results of the gravitational redshift, we first derive the exact null equations of motion in the RN geometry in harmonic coordinates. For a given light observer, a new numerical technique is proposed in the integration of the geodesic equations to determine the spatial position of the emitter,
considering the fact that their spatial positions are not always known simultaneously. Our work might be helpful to the related observations for probing strong gravity.
\end{abstract}

\section{Introduction}
The study on gravitational frequency shift of light signals is a important topic in modern astronomy since its original prediction by general relativity (GR)~\cite{Einstein11}. Acting as a crucial experimental verification of the local position invariance~\cite{Will2014} of Einstein equivalence principle, null gravitational shift (especially redshift) has been applied extensively from ground-based~\cite{PR1959,PR1960,PS1964,TWFMV1983,GNT1995,CP2014} to space-borne-based~\cite{VL1979,Vessot1980,KAC1990} measurements and has received its high-accuracy consistency (at the $7\times10^{-5}$ level) with GR's prediction. In contrast to these efforts devoted to the weak-field counterparts in the solar system, investigations of gravitational shift of spectrum lines produced in the strong-field region of a compact object (e.g., a neutron star or black hole) beyond the solar system were also performed with a fascinating prospect (see~\cite{Paradijs1979,Lindblom1984,CPM2002,SPZT2002,DP2003,Stairs2003,MW2006,PF2013,DS2015}, and references therein). In particular, M\"{u}ller and Wold~\cite{MW2006} adopted the ray-tracing technique to numerically investigate the gravitational redshift effect of emission lines from both the weak- and strong-field regions of a Kerr black hole, which was compared with the observations of galaxy Mrk 110~\cite{K2003}. Actually, probing strong gravitational fields using the gravitational shift tests along with other tests (e.g., Refs.~\cite{TWDW1992,Lyne2004,Valtonen2008,VE2000,KP2005,VK2008,Virbhadra2009,LWKCL2012,MM2014,YS2016}) plays a significant role in the examinations of general relativity as well as alternative theories of gravity~\cite{Will2014,KBCLSJ2004,Psaltis2008}. Two reasons are responsible for this. The first one is that strong-field regions in our Universe are common, however, most of actual tests have been restricted in our solar system~\cite{Psaltis2008}. It is necessary and naturally desired to grasp the degree to which the consistency of the predictions of gravitational theories including GR with the astronomical observations can achieve. A second one is that gravitational shift tests are classical and relative feasible, and they have become one of the most convenient chooses~\cite{DP2003}. Recently, the most direct evidence of a binary black hole system in the first direct detections of gravitational waves was reported~\cite{Abbott2016}. Very recently, the images of a supermassive black hole (i.e., the one at the center of M87 Galaxy) observed directly by the Event Horizon Telescope have also been presented for the first time~\cite{A2019,DD2019}. There is no doubt that these two events as well as other observations (e.g, Ref.~\cite{Abbott2016b}) may promote further the theoretical studies of gravitational shifts of frequency of light in the strong field of a black hole.

Historically, considerations of gravitational shift in the strong field limit were usually done under the common scenario, where the light receiver was far away from the central body while the light emitter was in the strong-field region of the lens, corresponding to the famous gravitational redshift phenomenon. Under this scenario, the gravitational shift approximately depends only on the emitter's position. However, further consideration is necessary if both the light emission and reception events happen in the strong-field region for probing strong gravity in future. This is because the gravitational shift for this case depends not only on the position of the receiver but also on that of the emitter, which, however, may not be always known simultaneously. This paper gives a new numerical approach to tackle the issue how to obtain the emitter's position (and thus the gravitational shift) for a given observer and a given received direction of the light curve.

In this work, based on the exact equations of motion of light in harmonic coordinates, we study the gravitational shift of light induced by a Reissner-Nordstr\"{o}m black hole numerically in the traditional strong-field scenario, as well as in the scenario where both the light emitter and the observer are located in the strong field of the lens. Since the emitter is connected to the receiver by a light-like world line, our focus is the determination of the unknown spatial position of one of the emitter and the receiver by integrating the null geodesics in the strong field limit, when the other's position is given. For this purpose, a new numerical technique is proposed. Two things should be pointed out. First, both the issue and the method of this work are different from that in the numerical investigation~\cite{MW2006} and theoretical considerations~\cite{PF2013,DS2015}. Second, we note that rapid progress in the precision to measure the gravitational shift of frequency has been made during the last two decades. For example, the gravitational frequency shift of radio photons was measured with a high precision ($\sim 10^{-14}$) in the Cassini experiment in 2002~\cite{IGAB1999,BIT2003,KPSV2007}. There might be a possibility to detect or constrain the intrinsic electric charge of a RN black hole by measuring its component contribution to the gravitational redshift, although it may be very weak up to now.

This paper is organized as follows. In Section~\ref{derivation}, the analytical form of the null gravitational redshift in Reissner-Nordstr\"{o}m spacetime in the strong field limit is presented in harmonic coordinates. In Section~\ref{Numerical simulations}, we first derive the exact null equations of motion in this spacetime, and then integrate them via a new numerical technique to calculate the spacial position of the light emitter for a given observer. The numerical results of the gravitational redshift are thus obtained for both the traditional and non-traditional strong-field scenarios. A summary is given in Section~\ref{Summary}. In what follows, we use natural units in which $G = c = 1$.

\section{Null Gravitational shift caused by a RN black hole in harmonic coordinates} \label{derivation}
\begin{figure*}[t]
\center
\begin{center}
  \includegraphics[width=14cm]{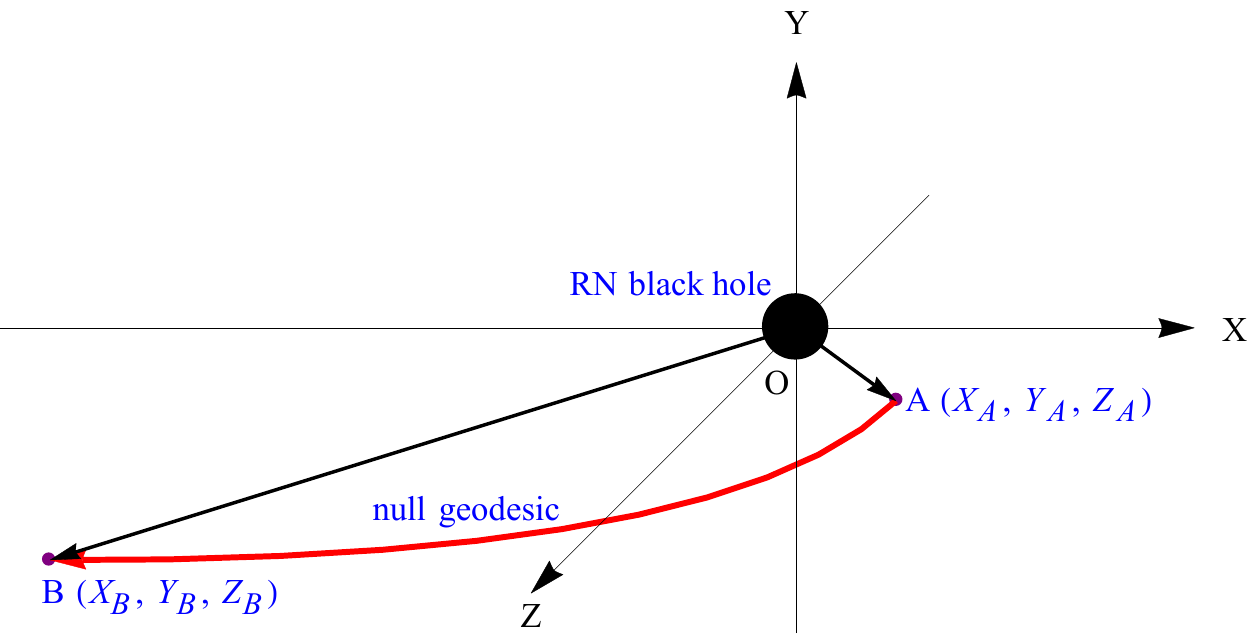}
  \caption{Schematic diagram for the light propagation in the gravitational field of a Reissner-Nordstr\"{o}m black hole. The central body is assumed to be located at the origin $O (0,~0,~0)$ of a three-dimensional Cartesian coordinate system $(X,~Y,~Z)$. }    \label{Figure1}
\end{center}
\end{figure*}
Consider the gravitational frequency shift of light signals in the Reissner-Nordstr\"{o}m geometry. The Reissner-Nordstr\"{o}m metric in harmonic coordinates $(X_0,~X_1,~X_2,~X_3)$ (defined as $(T,~X,~Y,~Z)$) can be written as~\cite{LinJiang2014}
\begin{eqnarray}
\nn&&ds^{2}=-\frac{R^2-M^2+Q^2}{(R+M)^2}dX_0^2+\left(1+\frac{M}{R}\right)^2   \\
&& \hspace*{28pt} \times\left(\delta_{ij}+\frac{M^2-Q^2}{R^2-M^2+Q^2}\frac{X_iX_j}{R^2}\right)dX_idX_j ~,~~~~~ \label{HRN}
\end{eqnarray}
where $\delta_{ij}$ denotes Kronecker delta, $i$ and $j$ run over the values $1,~2,~3$, $\bm{X}\cdot d\bm{X}\equiv X_1dX_1+X_2dX_2+X_3dX_3$, and $R\equiv|\bm{X}|=\sqrt{X_1^2+X_2^2+X_3^2}$. $M$ and $Q$ are the rest mass and electrical charge of the gravitational source, respectively, with the relation $M^2\geq Q^2$ to avoid the naked singularity of the black hole. Figure~\ref{Figure1} shows the schematic diagram for the propagation of light in the gravitational field of a RN black hole. The spacial coordinates of the light emitter (marked by $A$) and the receiver (marked by $B$) are denoted as $(X_A,~Y_A,~Z_A)$ and $(X_B,~Y_B,~Z_B)$, respectively. $A$ is connected to $B$ by a null world line (red line). We denote the coordinate times of the emission and reception events as $T_A$ and $T_B$ respectively. For simplicity, we discuss the special case in which both the light emitter and the receiver are assumed to be static in the background's rest frame $(T,~X,~Y,~Z)$.

The gravitational redshift of frequency of light is conventionally defined by~\cite{MW2006,Harrison1974,PB1993,KL2010}
\begin{eqnarray}
z\equiv\frac{\nu_A-\nu_B}{\nu_B}=\frac{d\tau_B}{d\tau_A}-1=\frac{dT_A}{d\tau_A}\frac{dT_B}{dT_A}\frac{d\tau_B}{dT_B}-1~, \label{definition}
\end{eqnarray}
where $\tau_A$ and $\tau_B$ denote the proper times of the light emitter and the receiver, respectively. A negative redshift means a blueshift. For the case of the emitter and the receiver being static, $\frac{d\tau_A}{dT_A}$, $\frac{dT_B}{d\tau_B}$, and $\frac{dT_A}{dT_B}$ in the Reissner-Nordstr\"{o}m spacetime read as follows:
\begin{eqnarray}
&&\frac{dT_A}{d\tau_A}=\frac{1}{\sqrt{-g_{00}\left(\bm{X}_A\right)}}=\sqrt{\frac{(R_A+M)^2}{R_A^2-M^2+Q^2}}~,  \label{FS1}   \\
&&\frac{d\tau_B}{dT_B}=\sqrt{-g_{00}\left(\bm{X}_B\right)}=\sqrt{\frac{R_B^2-M^2+Q^2}{(R_B+M)^2}}~,  \label{FS2} \\
&&\frac{dT_B}{dT_A}=1~,  \label{FS3}
\end{eqnarray}
where $R_A=\sqrt{X_A^2\!+\!Y_A^2\!+\!Z_A^2}$ and $R_B=\sqrt{X_B^2\!+\!Y_B^2\!+\!Z_B^2}$. Thus, the exact form of the gravitational redshift of frequency induced by a RN source in harmonic coordinates is easily rewritten as follows:
\begin{eqnarray}
z_{RN}=\sqrt{\frac{(R_B^2-M^2+Q^2)(R_A+M)^2}{(R_A^2-M^2+Q^2)(R_B+M)^2}}-1~. \label{FS-RN}
\end{eqnarray}

In the conventional strong-field case, the light observer is far away from the central body ($R_B\rightarrow+\infty$), and the redshift in Equation~(\ref{FS-RN}) reduces to $z_{RN}=\sqrt{\frac{(R_A+M)^2}{R_A^2-M^2+Q^2}}-1$. However, this is not the case in our non-traditional scenario where both the light emitter and the observer are located in the strong-field region. Notice that the positions of the receiver and the emitter may not be always known at the same time in some cases, which is different from the assumption that they are always known simultaneously given in previous work (see~\cite{SM2007,Hees2012,HBL2014}, and references therein). Since they are connected by a light-like geodesic, our focus is the determination of the position of one of the light emitter and the observer via integrating the exact null geodesics, once the spacial position of the other is given. Without loss of generality, the observer's position is assumed to be known while the emitter's position unknown in the following calculations.

\section{Numerical calculations in the strong field limit} \label{Numerical simulations}
In this section, we first derive the exact null equations of motion in the RN spacetime. A new numerical technique, which is different from the \emph{Kerr Black Hole Ray Tracer} technique~\cite{MW2006}, is then adopted to compute the gravitational redshift $z_{RN}$ in the strong field of the RN black hole for both the traditional and non-traditional strong-field scenarios.

\subsection{Exact null geodesics in the RN geometry} \label{NullGeodesics}
Based on the harmonic metric given in Equation~(\ref{HRN}), we obtain the exact form of the equations of motion of light in the RN geometry via calculating the nonvanishing Christoffel symbols tediously but straightforwardly as follows:
\begin{eqnarray}
0=\ddot{T}+\frac{2(M^2-Q^2+MR)(X\dot{X}+Y\dot{Y}+Z\dot{Z})\dot{T}}{R(R+M)(R^2-M^2+Q^2)} ~,~~~~~ \label{geodesic-t-s}
\end{eqnarray}
\begin{eqnarray}
\nn&&0=\ddot{X}+\frac{X(R^2-M^2+Q^2)(M^2-Q^2+MR)\dot{T}^2}{R(R+M)^5}    \\
\nn&&-\frac{2M(X\dot{X}+Y\dot{Y}+Z\dot{Z})\dot{X}}{R^2(M+R)}+\frac{X}{R^5(M+R)}    \\
\nn&&\times\Bigg{[}R^2(M^2-Q^2+MR)(\dot{X}^2+\dot{Y}^2+\dot{Z}^2)    \\
&&+\frac{(M^2\!\!-\!Q^2)(M^2\!\!-\!Q^2\!-\!MR\!-\!2R^2)(X\dot{X}\!+\!Y\dot{Y}\!+\!Z\dot{Z})^2}{R^2-M^2+Q^2} \Bigg{]} , \label{geodesic-x-s} \\
\nn&&0=\ddot{Y}+\frac{Y(R^2-M^2+Q^2)(M^2-Q^2+MR)\dot{T}^2}{R(R+M)^5}    \\
\nn&&-\frac{2M(X\dot{X}+Y\dot{Y}+Z\dot{Z})\dot{Y}}{R^2(M+R)}+\frac{Y}{R^5(M+R)}    \\
\nn&&\times\Bigg{[}R^2(M^2-Q^2+MR)(\dot{X}^2+\dot{Y}^2+\dot{Z}^2)    \\
&&+\frac{(M^2\!\!-\!Q^2)(M^2\!\!-\!Q^2\!-\!MR\!-\!2R^2)(X\dot{X}\!+\!Y\dot{Y}\!+\!Z\dot{Z})^2}{R^2-M^2+Q^2} \Bigg{]} , \label{geodesic-y-s}  \\
\nn&&0=\ddot{Z}+\frac{Z(R^2-M^2+Q^2)(M^2-Q^2+MR)\dot{T}^2}{R(R+M)^5}    \\
\nn&&-\frac{2M(X\dot{X}+Y\dot{Y}+Z\dot{Z})\dot{Z}}{R^2(M+R)}+\frac{Z}{R^5(M+R)}    \\
\nn&&\times\Bigg{[}R^2(M^2-Q^2+MR)(\dot{X}^2+\dot{Y}^2+\dot{Z}^2)    \\
&&+\frac{(M^2\!\!-\!Q^2)(M^2\!\!-\!Q^2\!-\!MR\!-\!2R^2)(X\dot{X}\!+\!Y\dot{Y}\!+\!Z\dot{Z})^2}{R^2-M^2+Q^2} \Bigg{]}, \label{geodesic-z-s}
\end{eqnarray}
where a dot denotes the derivative with respect to the affine parameter $\xi$ along the geodesic~\cite{Weinberg1972,WS2004}. It can be seen that, dropping the electrical charge $(Q=0)$ in Equations~(\ref{geodesic-t-s}) - (\ref{geodesic-z-s}), we can get the exact null geodesics of the Schwarzschild spacetime in harmonic coordinates
\begin{eqnarray}
&&0=\ddot{T}+\frac{2M(X\dot{X}+Y\dot{Y}+Z\dot{Z})\dot{T}}{R(R^2-M^2)} ~, \label{geodesic-t-rn}  \\
\nn&&0=\ddot{X}+\frac{MX(R-M)\dot{T}^2}{R(R+M)^3}-\frac{2M(X\dot{X}+Y\dot{Y}+Z\dot{Z})\dot{X}}{R^2(R+M)}  \\
&&+\frac{MX\left[R^2(\dot{X}^2+\dot{Y}^2+\dot{Z}^2)+\frac{M(M-2R)(X\dot{X}+Y\dot{Y}+Z\dot{Z})^2}{R^2-M^2}\right]}{R^5}~,~~~~ \label{geodesic-x-rn}  \\
\nn&&0=\ddot{Y}+\frac{MY(R-M)\dot{T}^2}{R(R+M)^3}-\frac{2M(X\dot{X}+Y\dot{Y}+Z\dot{Z})\dot{Y}}{R^2(R+M)}  \\
&&+\frac{MY\left[R^2(\dot{X}^2+\dot{Y}^2+\dot{Z}^2)+\frac{M(M-2R)(X\dot{X}+Y\dot{Y}+Z\dot{Z})^2}{R^2-M^2}\right]}{R^5}~,~~~~  \label{geodesic-y-rn}  \\
\nn&&0=\ddot{Z}+\frac{MZ(R-M)\dot{T}^2}{R(R+M)^3}-\frac{2M(X\dot{X}+Y\dot{Y}+Z\dot{Z})\dot{Z}}{R^2(R+M)}  \\
&&+\frac{MZ\left[R^2(\dot{X}^2+\dot{Y}^2+\dot{Z}^2)+\frac{M(M-2R)(X\dot{X}+Y\dot{Y}+Z\dot{Z})^2}{R^2-M^2}\right]}{R^5}~,~~~~  \label{geodesic-z-rn}
\end{eqnarray}
which can be reduced to the result for the equatorial propagation~\cite{LinHe2016} in the second post-Minkowskian approximation.

\subsection{Basics of numerical simulations} \label{Basics}
For the convenience of discussion, three assumptions are made. First, we follow Wucknitz and Sperhake's idea~\cite{WS2004} and assume the trajectory parameter $\xi$ to take the dimension of length. Second, since the receiver $B\,(X_B,~Y_B,~Z_B)$ for the traditional scenario is far enough away from the gravitational source, we assume $|X_B|$ to be large enough, with $X_B\ll Y_B<0$ and $X_B\ll Z_B<0$. We finally assume the light propagating vector $\bm{k}$ at the position of the receiver $B$ to be approximately parallel to the $X$-axis of the coordinate system, which is not difficult to be realized by adjusting the coordinate frame. Notice that the light emitter $A\,(X_A,~Y_A,~Z_A)$ is close to the black hole but beyond its gravitational radius, and that the receiver for the non-traditional scenario is denoted by $\bar{B}\,(X_{\bar{B}},~Y_{\bar{B}},~Z_{\bar{B}})$ and is located at some place (between $A$ and $B$) of the same null world line.

With the consideration of the reversibility of light paths, there are three steps to determine the spatial coordinates of the light emitter $A\,$ for calculating the redshift $z_{RN}$:

First, for convenience, we make our numerical integrating direction be reverse to the direction of the light propagation shown in Figure~\ref{Figure1}, namely, computing from the observer $B\,(X_B,~Y_B,~Z_B)$ to the emitter $A\,(X_A,~Y_A,~Z_A)$.

Second, for a given observer $B$ and a given $X_A$, we need to integrate Equations~(\ref{geodesic-t-s}) - (\ref{geodesic-z-s}) numerically to obtain $Y_A$ and $Z_A$. Notice that for a given $X_B$ in the starting conditions, the adjustment of the coordinates $Y_B$ and $Z_B$ will make us get any desired position for the emitter $A$. Since there is only one specific value of $\xi$ corresponding to one point of the photon's world line, we can obtain the numerical value of the affine parameter $\xi_A$ for the emitter via performing a definite integral for $\dot{X}$ over $\xi$ as follows:
\begin{eqnarray}
X_A-X_B=\int_{\xi_B}^{\xi_A} \dot{X}d\xi ~~~(>0)~, \label{X}
\end{eqnarray}
with the affine parameter $\xi_B$ of the observer being given to be $X_B$ in the asymptotically flat spacetime~\cite{WS2004,LinHe2017}. It is worth to point out that there might be several numerical values of $\xi_A$ to correspond to a given $X_A$ in the strong field limit, and we take the one whose value is closet to $X_A$ in the following calculations for physical reasons. After the determination of $\xi_A$, we can immediately get the numerical values of $Y_A$ and $Z_A$ from the following definite integrals:
\begin{eqnarray}
&&Y_A-Y_B=\int_{\xi_B}^{\xi_A} \dot{Y}d\xi ~, \label{Y}  \\
&&Z_A-Z_B=\int_{\xi_B}^{\xi_A} \dot{Z}d\xi ~. \label{Z}
\end{eqnarray}

Finally, knowing the desired spatial position of the light emitter $A(X_A,~Y_A,~Z_A)$, the gravitational redshift of light propagating from the emitter $A$ to the receiver $B$ in the field of a given RN black hole can be calculated according to Equation~(\ref{FS-RN}). We emphasize that the position of the observer $B$ is the place where the starting conditions are given for both the traditional and non-traditional strong-field cases,
and that the numerical process above is also applied to calculating the gravitational redshift of light propagating from the emitter $A$ to the receiver $\bar{B}$ caused by the same central body.

In the numerical calculations, the computation domain is set as $\xi\in[-X_{max},~X_{max}]$, with $X_{max}=-X_B~(\gg\sqrt{Y_{B}^2+Z_{B}^2})$, and we use $X_{max}$ to replace the infinity $+\infty$ for a large enough $X_{max}$. The initial and boundary conditions for both the traditional and non-traditional strong-field cases are given as follows:
\begin{eqnarray}
&&\left.\frac{dT}{d\xi}\right|_{\xi\rightarrow -\infty}\left(\approx \left.\frac{dT}{d\xi}\right|_{\xi\rightarrow -X_{max}}\right)=1~, \label{Initial-dotT} \\
&&\left.\frac{dX}{d\xi}\right|_{\xi\rightarrow -\infty}\left(\approx\left.\frac{dX}{d\xi}\right|_{\xi\rightarrow -X_{max}}\right)=1~,    \label{Initial-dotX} \\
&&\left.\frac{dY}{d\xi}\right|_{\xi\rightarrow -\infty}\left(\approx\left.\frac{dY}{d\xi}\right|_{\xi\rightarrow -X_{max}}\right)=0~,  \label{Initial-dotY} \\
&&\left.\frac{dZ}{d\xi}\right|_{\xi\rightarrow -\infty}\left(\approx\left.\frac{dZ}{d\xi}\right|_{\xi\rightarrow -X_{max}}\right)=0~,    \label{Initial-dotZ} \\
&&\left.T\right|_{\xi\rightarrow -\infty}\left(\approx \left.T\right|_{\xi\rightarrow -X_{max}}\right)=-X_{max}~, \label{Initial-T} \\
&&\left.X\right|_{\xi\rightarrow -\infty}\left(\approx \left.X\right|_{\xi\rightarrow -X_{max}}\right)=-X_{max}~,  \label{Initial-X} \\
&&\left.Y\right|_{\xi\rightarrow -\infty}\left(\approx\left.Y\right|_{\xi\rightarrow -X_{max}}\right)=Y_{B}~,    \label{Initial-Y} \\
&&\left.Z\right|_{\xi\rightarrow -\infty}\left(\approx\left.Z\right|_{\xi\rightarrow -X_{max}}\right)=Z_{B}~.    \label{Initial-Z}
\end{eqnarray}

\subsection{Numerical results} \label{Examples}

\subsubsection{The traditional case}  \label{Examples-A}
\begin{table*}
\centering
\caption{The numerical values of the redshift $z_{RN}$ for different $Y_B$, $Z_B$, and $X_A$ (in units of $M$). As an example, we set $X_B=-1.0\times10^{10}$ and $Q=0.05$. }   \label{Table1}
\begin{minipage}{14cm}
\begin{tabular*}{\textwidth}{@{\extracolsep{\fill}}lrrrrrl@{}}
\hline
$X_A$ & \multicolumn{1}{c}{$-100$} & \multicolumn{1}{c}{$-10$} & \multicolumn{1}{c}{$-5$} & \multicolumn{1}{c}{$-1$} & \multicolumn{1}{c}{$0$} & \multicolumn{1}{c}{$1$}  \\
$(Y_B,~Z_B)$ \\
\hline
        ($-$10,~$-$8)         & 0.00997  &  0.0666  &  0.0832   & 0.0950 & 0.0969 &  0.0981     \\
        ($-$5,~$-$5)          & 0.0100   &  0.0877  &  0.136    & 0.201  & 0.218  &  0.232      \\
        ($-$3.724,~$-$3.724)  & 0.0100   &  0.0946  &  0.162    & 0.301  & 0.360  &  0.431      \\
        ($-$3.673,~$-$3.673)  & 0.0100   &  0.0948  &  0.163    & 0.307  & 0.369  &  0.448      \\
\hline
\end{tabular*}
\end{minipage}
\end{table*}

\begin{figure*}[t]
\center
\begin{center}
  \includegraphics[width=12cm]{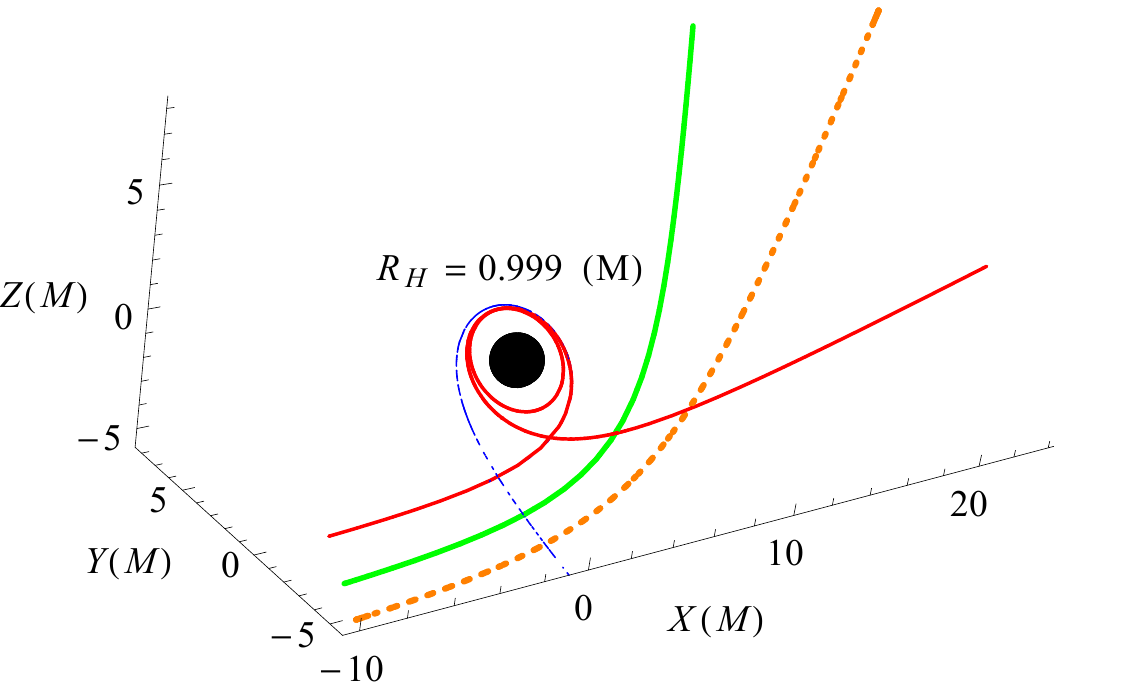}
  \caption{The three-dimensional null trajectories in the strong field of a RN black hole, plotted with different $Y_B$ and $Z_B$ (in units of $M$). Here, we assume $X_B=-1.0\times10^{10}$, $Q=0.05$, and $Y_B=Z_B=-6$ (dotted orange), $-5$ (thick green), $-3.6865$ (dot-dashed blue), and $-3.67271$ (thin red) as examples.  }    \label{Figure2}
\end{center}
\end{figure*}

We first consider the redshift $z_{RN}$ as the function of the positions of the emitter and the receiver. Since the event horizon of a RN black hole is located at $R_H=\sqrt{M^2-Q^2}$ in harmonic coordinates, the numerical values of $Y_B$ and $Z_B$ can not be too small for avoiding the swallow of light by the black hole. Table~\ref{Table1} gives the numerical values of the redshift $z_{RN}$ for variable $Y_B$, $Z_B$, and $X_A$ in units of $M$.  We can see that the magnitues of $z_{RN}$ are much larger than nowadays precesion ($\sim10^{-14}$) of the astronomical measurements of the shift. For the convenience of display, Figure~\ref{Figure2} shows the propagating trajectories of light in the strong field of a RN black hole, plotted for different $Y_B$ and $Z_B$.

We next discuss the contribution of the electrical charge $Q$ to the frequency shift. Table~\ref{Table2} gives the numerical values of the redshift $z_{RN}$ for variable $Q$ and $X_A$ in units of $M$. It indicates that it is possible to detect the charge-produced contribution to the gravitational redshift in the high-accuracy observations, although the electrical charge of the black hole may be weak (such as $Q=0.01M$).
\begin{table*}
\centering
\caption{The numerical values of the redshift $z_{RN}$ for different $Q$ (in units of $M$), with $X_B=-1.0\times10^{10}$, $Y_B=-10$, and $Z_B=-5$. } \label{Table2}
\begin{minipage}{14cm}
\begin{tabular*}{\textwidth}{@{\extracolsep{\fill}}lrrrrrl@{}}
\hline
$X_A$ & \multicolumn{1}{c}{$-100$} & \multicolumn{1}{c}{$-10$} & \multicolumn{1}{c}{$-5$} & \multicolumn{1}{c}{$-1$} & \multicolumn{1}{c}{$0$} & \multicolumn{1}{c}{$1$}  \\
$Q$   \\
\hline
        0.99                 & 0.009940278 & 0.06958253 & 0.08979824 & 0.1054559 & 0.1080660 & 0.1097624     \\
        0.1                  & 0.009988676 & 0.07211171 & 0.09416427 & 0.1118451 & 0.1149210 & 0.1170061     \\
        0.01                 & 0.009989170 & 0.07213762 & 0.09420916 & 0.1119113 & 0.1149921 & 0.1170815     \\
        0                    & 0.009989174 & 0.07213788 & 0.09420962 & 0.1119120 & 0.1149929 & 0.1170823     \\
\hline
\end{tabular*}
\end{minipage}
\end{table*}

\begin{table*}
\centering
\caption{The numerical values of the redshift $z_{RN}$ for different $M$ and $X_A$ (in units of $M_\odot$), with $X_B=-1.0\times10^{10}$, $Y_B=-1000$, $Z_B=-500$, and $Q=0.05M$. }   \label{Table3}
\begin{minipage}{14cm}
\begin{tabular*}{\textwidth}{@{\extracolsep{\fill}}lrrrrrl@{}}
\hline
$X_A$ & \multicolumn{1}{c}{$-100M$} & \multicolumn{1}{c}{$-10M$} & \multicolumn{1}{c}{$-5M$} & \multicolumn{1}{c}{$-M$} & \multicolumn{1}{c}{$0$} & \multicolumn{1}{c}{$M$}  \\
$M$   \\
\hline
        10                  & 0.00672 & 0.00910 & 0.00913 & 0.00915 & 0.00915 &  0.00915     \\
        50                  & 0.00981 & 0.0438  & 0.0479  & 0.0500  & 0.0503  &  0.0505      \\
        100                 & 0.00999 & 0.0721  & 0.0942  & 0.112   & 0.115   &  0.117       \\
        180                 & 0.0100  & 0.0911  & 0.147   & 0.239   & 0.268   &  0.295       \\
        200                 & 0.0100  & 0.0934  & 0.157   & 0.277   & 0.322   &  0.371       \\
\hline
\end{tabular*}
\end{minipage}
\end{table*}
With respect to the gravitational redshift as a function of the rest mass $M$ of the gravitational lens for a given observer, Table~\ref{Table3} presents the numerical values of $z_{RN}$ for different rest mass $M$ and the emitter's position ($X_A$) in units of $M_{\odot}$.

\begin{table*}
\centering
\caption{The numerical values of $z_{RN}$ for light propagating from the emitter $A$ to the receiver $\bar{B}$ for different $Y_B$, $Z_B$, and $X_A$ (in units of $M$).  }   \label{Table4}
\begin{minipage}{14cm}
\begin{tabular*}{\textwidth}{@{\extracolsep{\fill}}lrrrrl@{}}
\hline
$X_A$ & \multicolumn{1}{c}{$-10$} & \multicolumn{1}{c}{$-5$} & \multicolumn{1}{c}{$-1$} & \multicolumn{1}{c}{$0$} & \multicolumn{1}{c}{$1$}  \\
$(Y_B,~Z_B)$   \\
\hline
        ($-$10,~$-$8)           & 0.0221  &   0.0379   &  0.0492  & 0.0511 & 0.0523     \\
        ($-$5,~$-$5)            & 0.0374  &   0.0830   &  0.145   & 0.161  & 0.175      \\
        ($-$3.724,~$-$3.724)    & 0.0427  &   0.107    &  0.239   & 0.295  & 0.363      \\
        ($-$3.673,~$-$3.673)    & 0.0429  &   0.108    &  0.245   & 0.304  & 0.379      \\
\hline
\end{tabular*}
\end{minipage}
\end{table*}
\subsubsection{The non-traditional case} \label{Examples-B}
The numerical results above are based on the scenario where the light receiver $B$ is located far enough from the black hole. We now consider the gravitational redshift of light coming from the emitter $A~(X_A,~Y_A,~Z_A)$ to the receiver $\bar{B}~(X_{\bar{B}},~Y_{\bar{B}},~Z_{\bar{B}})$ ($X_B\ll X_{\bar{B}}<X_A$). With the starting conditions (i.e., Equations~(\ref{Initial-dotT}) - (\ref{Initial-Z})), the spacial positions of both the light emitter $A$ and the receiver $\bar{B}$ can be determined via the process in Subsection~\ref{Examples-A}, if $X_A$ and $X_{\bar{B}}$ are given. Note that in the scenario where both the light emitter and the receiver are located in the strong-field region, the spacial point $(X_B,~Y_B,~Z_B)$ is just regarded as the position of the starting conditions rather than a reception point. In units of $M$, Table~\ref{Table4} shows the numerical values of the gravitational redshift $z_{RN}$ of light propagating from $A$ to $\bar{B}$ for different values of $Y_B$, $Z_B$, and $X_A$, with $X_B=-1.0\times10^{10}$ and $X_{\bar{B}}=20$ as an example.

\section{Summary} \label{Summary}
In this work, the gravitational redshift effect of light in the Reissner-Nordstr\"{o}m geometry has been studied numerically, in both the traditional strong-field scenario and the non-traditional scenario where both the light emission and reception events happen in the strong-field region. In order to obtain the numerical results of the gravitational redshift, the exact null geodesics in the Reissner-Nordstr\"{o}m spacetime have been derived and solved via a new numerical technique to determine the spatial position of the emitter for a given light observer. Since the positions of the emission and reception events are not always known simultaneously, this work might be helpful to the related astronomical measurements for probing strong gravity in future.

\section*{ACKNOWLEDGEMENT}
This work was supported by the National Natural Science Foundation of China (Grant Nos. 11626129, 11801263, 11947018, and 11947128) and the Natural Science Foundation of Hunan (Grant No. 2018JJ3418).


\begin{thebibliography}{99}

 \bibitem{Einstein11} A. Einstein, Ann. Phys. {\bf 35}, 898 (1911)
 \bibitem{Will2014} C.M. Will, Living Rev. Relativ. {\bf 17}, 4 (2014)
 \bibitem{PR1959} R.V. Pound and G.A. Rebka, Jr., Phys. Rev. Lett. {\bf 3}, 439 (1959)
 \bibitem{PR1960} R.V. Pound and G.A. Rebka, Jr., Phys. Rev. Lett. {\bf 4}, 337 (1960)
 \bibitem{PS1964} R.V. Pound and J.L. Snider, Phys. Rev. Lett. {\bf 13}, 539 (1964)
 \bibitem{TWFMV1983} J.P. Turneaure, C.M. Will, B.F. Farrell, E.M. Mattison, and R.F.C. Vessot, Phys. Rev. D {\bf 27}, 1705 (1983)
 \bibitem{GNT1995} A. Godone, C. Novero, and P. Tavella, Phys. Rev. D {\bf 51}, 319 (1995)
 \bibitem{CP2014} T.R. Cort\'{e}s and P.L. Pall\'{e}, Mon. Not. R. Astron. Soc. {\bf 443}, 1837 (2014)

 \bibitem{VL1979} R.F.C. Vessot and M.W. Levine, Gen. Relat. Gravit. {\bf 10}, 181 (1979)
 \bibitem{Vessot1980} R.F.C. Vessot et al., Phys. Rev. Lett. {\bf 45}, 2081 (1980)
 \bibitem{KAC1990} T.P. Krisher, J.D. Anderson, and J.K. Campbell, Phys. Rev. Lett. {\bf 64}, 1322 (1990)
 \bibitem{Paradijs1979} J. van Paradijs, Astrophys. J. {\bf 234}, 609 (1979)
 \bibitem{Lindblom1984} L. Lindblom, Astrophys. J. {\bf 278}, 364 (1984)
 \bibitem{CPM2002} J. Cottam, F. Paerels, and M. Mendez, Nature {\bf 420}, 51 (2002)
 \bibitem{SPZT2002} D. Sanwal, G.G. Pavlov, V.E. Zavlin, and M.A. Teter, Astrophys. J. {\bf 574}, L61 (2002)
 \bibitem{DP2003} S. DeDeo and D. Psaltis, Phys. Rev. Lett. {\bf 90}, 141101 (2003)
 \bibitem{MW2006} A. M\"{u}ller and M. Wold, Astron. Astrophys. {\bf 457}, 485 (2006)
 \bibitem{Stairs2003} I.H. Stairs, Living Rev. Relativ. {\bf 6}, 5 (2003)
 \bibitem{PF2013} F. Payandeh and M. Fathi, Int. J. Theor. Phys. {\bf 52}, 3313 (2013)
 \bibitem{DS2015} A.K. Dubey and A.K. Sen, Astrophys. Space Sci. {\bf 360}, 29 (2015)

 \bibitem{K2003} W. Kollatschny, Astron. Astrophys. {\bf 412}, L61 (2003)
 \bibitem{TWDW1992} J.H. Taylor, A. Wolszzan, T. Damour, and J.M. Weisberg, Nature {\bf 355}, 132 (1992)
 \bibitem{Lyne2004} A.G. Lyne et al., Science {\bf 303}, 1153 (2004)
 \bibitem{Valtonen2008} M.J. Valtonen et al., Nature {\bf 452}, 851 (2008)
 \bibitem{VE2000} K.S. Virbhadra and G.F.R. Ellis, Phys. Rev. D {\bf 62}, 084003 (2000)
 \bibitem{KP2005} C.R. Keeton and A.O. Petters, Phys. Rev. D {\bf 72}, 104006 (2005)
 \bibitem{VK2008} K.S. Virbhadra and C.R. Keeton, Phys. Rev. D {\bf 77}, 124014 (2008)
 \bibitem{Virbhadra2009} K.S. Virbhadra, Phys. Rev. D {\bf 79}, 083004 (2009)
 \bibitem{LWKCL2012} K. Liu, N. Wex, M. Kramer, J.M. Cordes, and T.J.W. Lazio, Astrophys. J. {\bf 747}, 1 (2012)
 \bibitem{MM2014} M.C. Miller and J.M. Miller, Phys. Rep. {\bf 548}, 1 (2014)
 \bibitem{YS2016} K. Yagi and L.C. Stein, Class. Quantum Grav. {\bf 33}, 054001 (2016)
 \bibitem{KBCLSJ2004} M. Kramer, D.C. Backer, J.M. Cordes, T.J.W. Lazio, B.W. Stappers, and S. Johnston, New Astron. Rev. {\bf 48}, 993 (2004)
 \bibitem{Psaltis2008} D. Psaltis, Living Rev. Relativ. {\bf 11}, 9 (2008)

 \bibitem{Abbott2016} B.P. Abbott et al., Phys. Rev. Lett. {\bf 116}, 061102 (2016)
 \bibitem{A2019} K. Akiyama et al., Astrophys. J. Lett. {\bf 875}, L4 (2019)
 \bibitem{DD2019} H. Davoudiasl and P.B. Denton, Phys. Rev. Lett. {\bf 123}, 021102 (2019)
 \bibitem{Abbott2016b} B.P. Abbott et al., Phys. Rev. Lett. {\bf 116}, 241103 (2016)
 \bibitem{IGAB1999} L. Iess, G. Giampieri, J.D. Anderson, and B. Bertotti, Class. Quantum Grav. {\bf 16}, 1487 (1999)
 \bibitem{BIT2003} B. Bertotti, L. Iess, and P. Tortora, Nature {\bf 425}, 374 (2003)
 \bibitem{KPSV2007} S.M. Kopeikin, A.G. Polnarev, G. Sch\"{a}fer, and I. Yu. Vlasov, Phys. Lett. A {\bf 367}, 276 (2007)

 \bibitem{LinJiang2014} W. Lin and C. Jiang, Phys. Rev. D {\bf 89}, 087502 (2014)
 \bibitem{Harrison1974} E.R. Harrison, Astrophys. J. {\bf 191}, L51 (1974)
 \bibitem{PB1993} T. Pyne and M. Birkinshaw, Astrophys. J. {\bf 415}, 459 (1993)
 \bibitem{KL2010} M. Killedar and G.F. Lewis, Mon. Not. R. Astron. Soc. {\bf 402}, 650 (2010)
 \bibitem{SM2007} A. San Miguel, Gen. Relat. Gravit. {\bf 39}, 2025 (2007)
 \bibitem{Hees2012} A. Hees et al., Class. Quantum Grav. {\bf 29}, 235027 (2012)
 \bibitem{HBL2014} A. Hees, S. Bertone, and C. Le Poncin-Lafitte, Phys. Rev. D {\bf 89}, 064045 (2014)
 \bibitem{Weinberg1972} S. Weinberg, \textit{Gravitation and Cosmology: Principles and Applications of the General Theory of Relativity} (Wiley, New York, 1972)
 \bibitem{WS2004} O. Wucknitz and U. Sperhake, Phys. Rev. D {\bf 69}, 063001 (2004)
 \bibitem{LinHe2016} G. He and W. Lin, Class. Quantum Grav. {\bf 33}, 095007 (2016)
 \bibitem{LinHe2017} G. He and W. Lin, Class. Quantum Grav. {\bf 34}, 105006 (2017)

\end{thebibliography}
\end{document}